\documentclass[english]{cccconf}
\usepackage[fleqn]{amsmath}
\usepackage[comma,numbers,square,sort&compress]{natbib}
\usepackage{subfigure}
\usepackage{ntheorem,amsmath,graphicx,float,caption,amssymb,flushend}
\usepackage{epstopdf}
\usepackage[justification=centering]{caption}
\usepackage{natbib}
\usepackage{makecell}
\usepackage{multirow}
\newtheorem{Theorem}{\textit{Theorem}}
\newtheorem*{Proof}{\textit{Proof}}
\newtheorem{Lemma}{\textit{Lemma}}
\newtheorem{Remark}{\textit{Remark}}
\newtheorem{Assumption}{\textit{Assumption}}
\newtheorem{Definition}{\textit{Definition}}
\newtheorem{Step}{\textit{Step}}

\begin{document}

\title{Event-Triggered Consensus for Linear Continuous-time Multi-agent Systems Based on a Predictor}
\author{Xiaoyu Liu, Jian Sun*, Lihua Dou, Jie Chen}



\affiliation{Key Laboratory of Intelligent Control and Design of Complex Systems \\Beijing Institute of Technology, Beijing, P.~R.~China, 100081
        \email{sunjian@bit.edu.cn}}

\maketitle

\begin{abstract}
In this paper, the problem of event-triggered consensus for linear continuous-time multi-agent systems is investigated. A new event-triggered consensus protocol based on a predictor is proposed to achieve consensus without continuous communication among agents. In the proposed consensus protocol, each agent only needs to monitor its states to determine its event-triggered instants. When an event is triggered, the agent will update its consensus protocol and sent its state information to its neighbors. In addition, the agent will also update its consensus protocol and the predictor when it receives the state information from its neighbors. A necessary and sufficient condition that the consensus problem can be solved is derived. Moreover, it is proved that Zeno behavior does not exist. Finally, a numerical example is given to illustrate that the protocol proposed in this paper can make the multi-agent systems achieve consensus through much fewer event-triggered times.
\end{abstract}

\keywords{Multi-agent systems, Event-triggered strategy, Consensus, Predictor }

\section{Introduction}
In the 1970s, the definition of agent was proposed in the field of intelligence\cite{ref1}. Then more and more researchers began to pay their attention to agents and rich results have been obtain. To mention a few, the consensus problem of multi-agent systems with the directed communication topology and the one-order integrator dynamics was investigated and the theoretical framework for the consensus problem of multi-agent systems was built in \cite{ref2}. The consensus problem of multi-agent systems with one-order integrator dynamics, active leaders and variable interconnection topologies was considered in \cite{ref3}. For the multi-agent systems with second-order integrator dynamics, an necessary and sufficient condition for the consensus was proposed in \cite{ref4}. The leader-following consensus problem of second-order nonlinear multi-agent systems with general topologies was studied without assuming that the interaction diagraph was strongly connected or contained a directed spanning tree in \cite{ref5}. And for the multi-agent system with high-order integrator dynamics, a necessary and sufficient condition was proposed for the consensus problem in \cite{ref6}. The consensus for high-order linear multi-agent systems with time delays in both the communication channel and control inputs was investigated in \cite{ref7}. The consensus problem of multi-agent systems with fixed/switching communication topology was investigated in \cite{ref8} using the Lyapunov method. The existence of consensus protocols for linear continuous-time/discrete-time multi-agent systems with fixed communication topology was proved in \cite{ref9} and \cite{ref10}. And other results about multi-agent systems can be seen in \cite{ref11,ref12} and references therein. 

It should be noticed that all the above publications assumed that there exists continuous communication between agents to implement the consensus protocol. However, it was well known that the continuous communication between agents was impossible in practice since the network bandwidth and the energy of agents were limited. And the continuous communication between agents would also result in the waste of communication resources \cite{ref13,ref14,ref15,ref16}. In order to avoid continuous communication and save the communication resources, the event-triggered strategy has received more and more attention. The consensus protocol was designed for multi-agent systems with the one-order integrator dynamics based on a self-triggered strategy in \cite{ref17}. Event-triggered consensus protocols were designed for multi-agent systems with the one-order/second-order integrator dynamics in \cite{ref18}. Two event-triggered consensus protocols were designed for multi-agent systems with the general linear dynamics in \cite{ref19}, but both the protocols were only effective for the undirected communication topology. For multi-agent systems with the general linear dynamics and the directed communication topology, the event-triggered consensus problem was investigated in \cite{ref20}. The consensus protocol in \cite{ref20} could make multi-agent systems achieve consensus without continuous communication, but the state differences between agents would merely converge to the neighbourhood of 0. In \cite{ref21} a distributed consensus protocol was designed to make the state differences between agents converge to 0 ultimately based on an event-triggered strategy and a necessary and sufficient condition was proposed for the consensus. 

In this paper, a new event-triggered consensus protocol is proposed for the multi-agent systems with general linear continuous-time dynamics based on a predictor. The communication topology among agents is assumed to be general directed. Under the consensus protocol and the triggering function proposed in this paper, the multi-agent systems can achieve consensus without continuous communication. Then, the Zeno behavior is proved to be nonexistent. In addition, the method proposed in this paper can make the multi-agent systems achieve consensus with much fewer event-triggering times than the existing methods. 

The rest of this paper is organized as follow. Some useful notations and the graph theory are introduced in Section 2. The design of the consensus protocol based on the event-triggered strategy is given in Section 3. In Section 4, the analysis of the consensus protocol is presented. A numerical example is given in Section 5 to illustrate the efficiency and the advantage of the event-triggered consensus protocol presented in this paper. At last, Section 6 concludes the paper.

\section{Notation and graph theory}
The notation and the graph theory used in this paper are introduced in this section. Let $\mathcal{R}^{m\times n}$ denote the set of $m\times n$ real matrices. $0_{m\times n}$ denotes the $m\times n$ matrix with all zeros. $I_{m\times n}$ and $I_{n}$ denote the $m\times n$ and $n\times n$ identity matrix respectively. $1_{n}$ denotes the $n\times 1$ column vector of all ones. A diagonal matrix with $x_{i}(i=1,2,\cdots,n)$ is denoted by $diag(x_{1},x_{2},\cdots,x_{n})$. $A\otimes B$ denotes the Kronecker product of matrices $A$ and $B$. Let $\Vert *\Vert$ denote the Euclidean norm for vectors and the induced 2-norm for matrices, respectively. $Re(*)$ denotes the real part of a complex number and $\lambda_{i}(*)$ denotes the $i$th eigenvalue of a matrix.

The communication topology among the $N$ agents is represented by a weighted graph $\mathcal{G}=(\mathcal{V},\varepsilon,\mathcal{A})$. $N$ agents in a multi-agent system are regarded as nodes $\mathcal{V}={1,2,\cdots,N}$ of the graph $\mathcal{G}$. A directed graph contains a directed spanning tree if there are directed paths from one node to every other ones. The adjacency matrix is defined as $\mathcal{A}=[a_{ij}]\in\mathcal{R}^{N\times N}$ associated with the directed graph $\mathcal{G}$. Assume that for all $i\in\ \mathcal{V}$, $a_{ii}=0$, $a_{ij}>0$ if $e_{ij}\in\ \varepsilon$ and $a_{ij}=0$ otherwise. The directed edge $e_{ij}\in\ \varepsilon$ denotes that agent $i$ can receive information from agent $j$. So agent $i$ can be called as agent $j$'s in-neighbor agent and agent $j$ can be called as agent $i$'s out-neighbor agent. $\mathcal{L}=[l_{ij}]\in \mathcal{R}^{N\times N}$ denotes the Laplacian matrix of the directed graph $\mathcal{G}$, where $l_{ii}=\sum_{j=1}^{N}\ a_{ij}$ and $l_{ij}=-a_{ij}(i\neq j)$.

\section{Design of the event-triggered consensus protocol}
A linear continuous-time multi-agent system is consisted of $N$ agents, where the dynamics of agent $i$ is described by
\begin{equation}
\label{1}
\dot{x}_{i}(t)=Ax_{i}(t)+Bu_{i}(t)
\end{equation}
where $x_{i}(t)\in\mathcal{R}^{n\times 1}$ and $u_{i}(t)\in\mathcal{R}^{m\times 1}$ are the state and the control input, respectively. $A\in\mathcal{R}^{n\times n}$, $B\in\mathcal{R}^{n\times m}$ are constant matrices. The communication topology among the $N$ agents can be described by a directed weighted graph $\mathcal{G}$. Assumption 1 is necessary to obtain the main result.
\begin{Assumption}
\label{A1}
The matrix pair $(A,B)$ in $\eqref{1}$ is stabilizable and the graph $\mathcal{G}$ contains a directed spanning tree.
\end{Assumption}
The well-known consensus protocol for the multi-agent system $\eqref{1}$ is
\begin{equation}
\label{2}
u_{i}(t)=K\sum_{j=1}^{N}\ a_{ij}\big[x_{i}(t)-x_{j}(t)\big]
\end{equation}
In order to apply the protocol \eqref{2}, a continuous communication between agent $i$ and $j$ is needed. For the purpose of saving the communication costs among agents, the event-triggered strategy is applied to design the consensus protocol. Under the event-triggered strategy, an event is designed for each agent in the multi-agent system. And the agent broadcasts its current information to its out-neighbor agents only when its event is triggered. The following consensus protocol is designed
\vspace{-2mm}
\begin{eqnarray}
\label{3}
\lefteqn{u_{i}(t)=K\sum_{j=1}^{N}\ a_{ij}\big[\hat{x}_{i}(t)-\hat{x}_{j}(t)\big]} & & \hspace{-8mm} \nonumber \\
&=& K\sum_{j=1}^{N}\ a_{ij}\big[\big(e^{A(t-t_{k_{i}}^{i})}x_{i}(t_{k_{i}}^{i}) \nonumber \\
& & +\int_{t_{k_{i}}^{i}}^{t}e^{A(t-s)}B\hat{u}_{i}(s)ds\big) \nonumber \\
& & -\big(e^{A(t-t_{k_{j}}^{j})}x_{j}(t_{k_{j}}^{j})+\int_{t_{k_{j}}^{j}}^{t}e^{A(t-s)}B\hat{u}_{j}(s)ds\big)\big]
\end{eqnarray}
where $K\in\mathcal{R}^{m\times n}$ is the feedback controlled gain matrix to be determined. $t_{k_{i}}^{i}$ is the most recent triggering instant of agent $i$, $k_{i}=1,2,3,\cdots$ represents the sequence number of the triggering instant of the agent $i$. $x_{i}(t_{k_{i}}^{i})$ is the last broadcast state of agent $i$. $\hat{x}_{i}(t)$ and $\hat{u}_{i}(t)$ represent the estimation of the state and the control input of the agent $i$, respectively. 

Then the measurement error is defined as
\begin{equation}
\label{4}
e_{i}(t)=e^{A(t-t_{k_{i}}^{i})}x_{i}(t_{k_{i}}^{i})+\int_{t_{k_{i}}^{i}}^{t}e^{A(t-s)}B\hat{u}_{i}(s)ds-x_{i}(t)
\end{equation}
And the triggering function is defined as
\begin{equation}
\label{5}
f_{i}(t)=\big\|e_{i}(t)\big\|-c_{1}e^{-\alpha t}
\end{equation}
where $c_{1}>0$, $0<\alpha<-maxRe(\lambda_{i}(\Pi))$ and $\Pi$ is defined after $\eqref{20}$.

From the triggering function $\eqref{5}$, it can be seen that when the triggering function $f_{i}(t)\geq 0$, agent $i$'s event is triggered. Then agent $i$ sends its current information including its state and state differences between agent $i$ and its in-neighbor agents to its out-neighbor agents and updates its consensus protocol. At the same time, the measurement error $e_{i}(t)$ is reset to 0. If the triggering function $f_{i}(t)<0$, it means that the communication from agent $i$ to its out-neighbor agents is unnecessary until the next event is triggered. On the other hand, agent $i$ will update its consensus protocol as soon as it receives the information from its in-neighbor agents. And the events of all agents are assumed to be triggered at the initial instant.

From $\eqref{3}$, it can be seen that the main challenge of the event-triggered consensus protocol proposed in this paper is how to obtain the estimation of the control input. Next, a method of estimating the control input is presented. 

Define $\theta_{ip}(t)=x_{i}(t)-x_{p}(t)$ and $\theta_{i}=[\theta^{T}_{i1}(t),\theta^{T}_{i2}(t), \\ \cdots,\theta^{T}_{i(i-1)}(t),\theta^{T}_{i(i+1)}(t),\cdots,\theta^{T}_{iN}(t)]^{T}$,
and $\eqref{2}$ can be rewritten as
\begin{equation}
\label{6}
u_{i}(t)=K(a^{*}_{i}\otimes I_{n})\theta_{i}(t)
\end{equation}
If applying the protocol \eqref{2} to system \eqref{1}, it is clear that
\begin{equation}
\begin{split}
\label{7}
\dot{\theta}_{ip}(t)=&\dot{x}_{i}(t)-\dot{x}_{p}(t) \\
=&A(x_{i}(t)-x_{p}(t))+BK(a^{*}_{i}\otimes I_{n})\theta_{i}(t)\\
&-BK(a^{*}_{p}\otimes I_{n})\theta_{p}(t)
\end{split}
\end{equation}
where $a^{*}_{i}=[a_{i1},a_{i2},\cdots,a_{i(i-1)},a_{i(i+1)},\cdots,a_{iN}]$.

Then $\eqref{7}$ can be rewritten as
\begin{equation}
\label{8}
\dot{\theta}_{i}(t)=\Omega_{i}\theta_{i}(t) \quad (i=1,2,\cdots,N)
\end{equation}
\vspace{-1mm}
where
\vspace{-3mm}
\begin{eqnarray*}
\Omega_{i}&=&I_{N-1}\otimes A+(d_{i}+1_{N-1}a_{i}^{*}-\mathcal{A}_{i}^{*})\otimes BK \\
d_{i}&=&diag(l_{11},l_{22},\cdots,l_{(i-1)(i-1)},l_{(i+1)(i+1)},\cdots,l_{NN}) \\
\mathcal{A}_{i}^{*}&= & \begin{bmatrix}\begin{smallmatrix}
 a_{11}&a_{12}&\cdots&a_{1(i-1)}&a_{1(i+1)}&\cdots&a_{1N} \\ a_{21}&a_{22}&\cdots&a_{2(i-1)}&a_{2(i+1)}&\cdots&a_{2N}\\ \vdots&\vdots&&\vdots&\vdots&&\vdots  \\ a_{(i-1)1}&a_{(i-1)2}&\cdots&a_{(i-1)(i-1)}&a_{(i-1)(i+1)}&\cdots&a_{(i-1)N} \\ a_{(i+1)1}&a_{(i+1)2}&\cdots&a_{(i+1)(i-1)}&a_{(i+1)(i+1)}&\cdots&a_{(i+1)N} \\ \vdots&\vdots&&\vdots&\vdots&&\vdots \\ a_{N1}&a_{N2}&\cdots&a_{N(i-1)}&a_{N(i+1)}&\cdots&a_{NN}
\end{smallmatrix}\end{bmatrix}
\end{eqnarray*}
And from $\eqref{6}$, it can be known that the estimation problem of agent $i$'s control input can be transformed into the estimation problem of the state differences between agent $i$ and its neighbor agents. On the basis of $\eqref{8}$, the following predictor is designed to estimate the state difference between agent $i$ and its neighbor agents.
\begin{equation}
\begin{split}
\label{9}
\hat{\theta}_{i}(t)=e^{\Omega_{i}(t-t_{k_{j}}^{j})}\hat{\theta}_{i}(t_{k_{j}}^{j}),  \quad  t\geq t_{k_{j}}^{j}
\end{split}
\end{equation}
where $\hat{\theta}_{i}(t^{j}_{k_{j}})=[\hat{\theta}^{T}_{i1}(t^{j}_{k_{j}}),\hat{\theta}^{T}_{i2}(t^{j}_{k_{j}}),\cdots,(x_{i}(t^{j}_{k_{j}})-x_{j}(t^{j}_{k_{j}}))^{T},\cdots,\hat{\theta}^{T}_{iN}(t^{j}_{k_{j}})]^{T}$, $t_{k_{j}}^{j}$ is the most recent triggering instant of agent $i$'s in-neighbor agent $j$.
\begin{Remark}
\label{r22}
It should be noted that \eqref{9} utilizes the \emph{artificial} closed-loop system \eqref{7} to predict the future state. Such a kind of predictor was first proposed in our previous work \cite{ref22a} and got further studied in \cite{ref22b,ref22c}.
\end{Remark}
From $\eqref{3}$ and $\eqref{9}$, it can be seen that if agent $j$ is triggered, then agent $j$ will send its current state $x_{j}(t_{k_{j}}^{j})$ and state difference $\hat{\theta}_{j}(t_{k_{j}}^{j})$ to agent $i$ at the triggering instant $t^{j}_{k_{j}}$. At the same time, agent $i$ updates the state difference between itself and agent $j$ using the state information $x_{j}(t^{j}_{k_{j}})$. So $\hat{\theta}_{i}(t) (t\geq t^{j}_{k_{j}})$ can be obtained based on the updated $\hat{\theta}_{i}(t^{j}_{k_{j}})$ and the triggering instant $t^{j}_{k_{j}}$. The estimation of the control input $\hat{u}_{i}(t)$ and $\hat{u}_{j}(t)$ in $\eqref{3}$ can be obtained
\begin{equation}
\begin{split}
\label{10}
\hat{u}_{i}(t)=K(a^{*}_{i}\otimes I_{n})e^{\Omega_{i}(t-t^{i}_{k_{i}})}\hat{\theta}_{i}(t^{i}_{k_{i}}),  \quad  t\geq t^{i}_{k_{i}}
\end{split}
\end{equation}
\vspace{-3mm}
\begin{equation}
\begin{split}
\label{11}
\hat{u}_{j}(t)=K(a^{*}_{j}\otimes I_{n})e^{\Omega_{j}(t-t^{j}_{k_{j}})}\hat{\theta}_{j}(t^{j}_{k_{j}}),  \quad  t\geq t^{j}_{k_{j}}
\end{split}
\end{equation}
where $t_{k_{i}}^{i}$ and $t_{k_{j}}^{j}$ are the most recent triggering instants of agent $i$ and agent $j$ respectively.
\begin{Definition}
\label{D1}
For the linear continuous-time multi-agent system $\eqref{1}$, if $\lim_{t\to\infty}\Arrowvert x_{i}(t)-x_{j}(t)\Arrowvert=0$ holds, it can be said that the protocol $\eqref{3}$ can solve the consensus problem or the multi-agent system $\eqref{1}$ can achieve consensus under the protocol $\eqref{3}$.
\end{Definition}
\begin{Lemma}
\label{L1}\cite{ref22}
If the graph $\mathcal{G}$ contains a directed spanning tree, zero is the simple eigenvalue of the Laplacian matrix $\mathcal{L}$ and all the other eigenvalues have positive real parts. Otherwise, $1_{N}$ is a right eigenvector associated with the zero eigenvalue.
\end{Lemma}
\begin{Lemma}
\label{L2}\cite{ref23}
For the Hurwitz matrix $M\in\mathcal{R}^{n\times n}$, when $t\geq 0$, there exist a $c_{M}>0$ such that $\big\|e^{Mt}\big\|\leq  c_{M}e^{\mu_{M}t}$ holds, where $max\{Re(\lambda_{i}(M))\}<\mu_{M}<0$.
\end{Lemma}
\begin{Lemma}
\label{L3}
For the linear continuous-time multi-agent system $\eqref{1}$ with the event-triggered consensus protocol $\eqref{3}$ and the triggering function $\eqref{5}$, if all the matrices $A+\lambda_{s}(\mathcal{L})BK$ $(s=2,3,\cdots,N)$ are Hurwitz, then all the matrices $\Omega_{i}=I_{N-1}\otimes A+(d_{i}+1_{N-1}a_{i}^{*}-\mathcal{A}_{i}^{*})\otimes BK$ $(i=1,2,\cdots,N)$ are also Hurwitz.
\end{Lemma}
\begin{Proof}
An invertible matrix can be taken as $S_{i}^{-1}=\left[\begin{array}{cc} P_{i}   \\  Q_{i}  \end{array} \right]$, where $P_{i}=[1,0,0,\cdots,0]\in R_{1\times N}$, $Q_{i}\in R_{(N-1)\times N}$ is a matrice which is derived by inserting $-1_{N-1}$ before the $i$th column or after the $i-1$th column of the identity matrix $I_{N-1}$, i.e.\\
\\
$Q_{i}=\left[\begin{array}{ccccccc} 1&0&\cdots&-1&\cdots&0&0 \\ 0&1&\cdots&-1&\cdots&0&0\\ \vdots&\vdots& &\vdots& &\vdots&\vdots  \\ 0&0&\cdots&-1&\cdots&1 &0\\ 0&0&\cdots&-1&\cdots&0 &1 \end{array} \right]$.
\\

By the definition of the Laplacian matrix $\mathcal{L}$, it is clear that
\begin{equation}
\begin{split}
\label{12}
S^{-1}_{i}\mathcal{L}S_{i}=\left[\begin{array}{cc} 0 & l_{1}^{i} \\  0 & d_{i}+1_{N-1}a_{i}^{*}-\mathcal{A}_{i}^{*} \end{array} \right]
\end{split}
\end{equation}
where $l_{1}^{i}=[l_{11},l_{12},\cdots,l_{1(i-1)},l_{1(i+1)},\cdots,l_{1N}]$.

It is assumed that $\lambda_{1}(\mathcal{L})=0,\lambda_{2}(\mathcal{L}),\cdots,\lambda_{N}(\mathcal{L})$ are the eigenvalues of the Laplacian matrix $\mathcal{L}$. From $\eqref{12}$, it can be seen that $\lambda_{s}(\mathcal{L})(s=2,3,\cdots,N)$ are the eigenvalues of $d_{i}+1_{N-1}a_{i}^{*}-\mathcal{A}_{i}^{*}$. Therefore, there exists an invertible matrix $T_{i}$ such that $d_{i}+1_{N-1}a_{i}^{*}-\mathcal{A}_{i}^{*}$ is similar to a Jordan canonical matrix.
\begin{equation}
\begin{split}
\label{13}
T^{-1}_{i}(d_{i}+1_{N-1}a_{i}^{*}-\mathcal{A}_{i}^{*})T_{i}=J_{i}=diag(J_{1}^{i},J_{2}^{i},\cdots,J_{m_{i}}^{i})
\end{split}
\end{equation}
where $J_{k}^{i} (k=1,2,\cdots,m_{i})$ are upper triangular Jordan blocks. And the principal diagonal elements of $J_{k}^{i}$ are $\lambda_{s}(\mathcal{L})(s=2,3,\cdots,N)$.

Therefore, the following equation can be obtained
\begin{equation}
\begin{split}
\label{14}
&(T_{i}\otimes I_{n})^{-1}(I_{N-1}\otimes A+(d_{i}+1_{N-1}a_{i}^{*}-\mathcal{A}_{i}^{*})\otimes BK)\times \\
&(T_{i}\otimes I_{n})=I_{N-1}\otimes A+J_{i}\otimes BK
\end{split}
\end{equation}
where $I_{N-1}\otimes A+J_{i}\otimes BK$ is an upper triangular block matrix.

According to the properties of Kronecker product\cite{ref24}, it can be known that the eigenvalues of $I_{N-1}\otimes A+J_{i}\otimes BK$ are given by the eigenvalues of $A+\lambda_{s}(\mathcal{L})BK(s=2,3,\cdots,N)$, i.e. the eigenvalues of the matrix $\Omega_{i}$ are the same as the ones of $A+\lambda_{s}(\mathcal{L})BK(s=2,3,\cdots,N)$. As a result, if all the matrices $A+\lambda_{s}(\mathcal{L})BK(s=2,3,\cdots,N)$ are Hurwitz, the matrice $\Omega_{i}$ is surely Hurwitz. The proof is completed. 
\end{Proof}

\section{Analysis of the event-triggered consensus protocol}

The following theorem presents the main results of this paper.
\begin{Theorem}
\label{T1}
Under the event-triggered consensus protocol $\eqref{3}$ and the triggering function $\eqref{5}$, the consensus problem of the linear continuous-time multi-agent system $\eqref{1}$ with a directed topology $\mathcal{G}$ can be solved without continuous communication if and only if all the matrices $A+\lambda_{i}(\mathcal{L})BK$ $(i=2,3,\cdots,N)$ are Hurwitz, where $\lambda_{i}(\mathcal{L})\neq 0$. In addition, the Zeno behavior does not exist.
\end{Theorem}
\begin{Proof}
(Sufficiency)
From the measurement error $\eqref{4}$, it is clear that
\begin{equation}
\label{15}
e^{A(t-t_{k_{i}}^{i})}x_{i}(t_{k_{i}}^{i})+\int_{t_{k_{i}}^{i}}^{t}e^{A(t-s)}B\hat{u}_{i}(s)ds=x_{i}(t)+e_{i}(t)
\end{equation}
Substituting $\eqref{15}$ into $\eqref{3}$ yields
\begin{equation}
\begin{split}
\label{16}
u_{i}(t)&=K\sum_{j=1}^{N}\ a_{ij}[x_{i}(t)+e_{i}(t)-x_{j}(t)-e_{j}(t)] \\
&=K[l_{i}x(t)+l_{i}e(t)]
\end{split}
\end{equation}
where $l_{i}=[l_{i1},l_{i2},\cdots,l_{iN}]$ represents the $i$th row of the Laplacian matrix $\mathcal{L}$, $x(t)=[x_{1}^{T}(t),x_{2}^{T}(t),\cdots,x_{N}^{T}(t)]^{T}$ and $e(t)=[e_{1}^{T}(t),e_{2}^{T}(t),\cdots,e_{N}^{T}(t)]^{T}$. \\
Then substituting $\eqref{16}$ into $\eqref{1}$ yields
\begin{equation}
\label{17}
\dot{x}_{i}(t)=Ax_{i}(t)+BK[l_{i}x(t)+l_{i}e(t)]
\end{equation}
Define $\delta_{i}(t)=x_{i}(t)-x_{1}(t)$, then it can be known that the multi-agent system $\eqref{1}$ will achieve consensus when $\lim_{t\to\infty}\Arrowvert \delta_{i}(t) \Arrowvert=0$ holds. On the basis of $\eqref{17}$, one can obtain that
\begin{align}
\label{18}
\dot{\delta}_{i}(t)&=\dot{x}_{i}(t)-\dot{x}_{1}(t)\nonumber\\
&=A\delta_{i}(t) \nonumber \\
&+BK\sum_{j=1}^{N}\ a_{ij}\big[x_{i}(t)+e_{i}(t)-x_{j}(t)-e_{j}(t)\big] \nonumber \\
&-BK\sum_{j=1}^{N}\ a_{1j}\big[x_{1}(t)+e_{1}(t)-x_{j}(t)-e_{j}(t)\big]
\end{align}
And $\eqref{18}$ can be transformed into the following form.
\begin{equation}
\begin{split}
\label{19}
\dot{\delta}(t)=&[I_{N-1}\otimes A+(\mathcal{L}_{22}+1_{N-1}a_{1}^{*})\otimes BK]\delta(t)\\
&+[(\mathcal{A}_{22}+1_{N-1}a_{1}+\mathcal{M})\otimes BK]e(t)
\end{split}
\end{equation}

\begin{align*}
 where \quad\delta(t)=&[\delta_{2}^{T}(t),\delta_{3}^{T}(t),\cdots,\delta_{N}^{T}(t)]^{T}, \\
e(t)=&[e_{1}^{T}(t),e_{2}^{T}(t),\cdots,e_{N}^{T}(t)]^{T}, \\
a_{1}^{*}=&[a_{12},a_{13},\cdots,a_{1N}], \\
a_{i}=&[a_{i1},a_{i2},\cdots,a_{iN}], \\
\mathcal{M}=&\left[\begin{array}{cccc} \l_{11}&0&\cdots&0 \\ l_{11}&0&\cdots&0\\ \vdots&\vdots& &\vdots  \\ l_{11}&0&\cdots&0  \end{array} \right]\in\mathcal{R}^{(N-1)\times N}
\end{align*}

\begin{align*}
\mathcal{L}_{22}=&\left[\begin{array}{cccc} l_{22}&-a_{23}&\cdots&-a_{2N} \\ -a_{32}&l_{33}&\cdots&-a_{3N}\\ \vdots&\vdots&\ddots &\vdots  \\ -a_{N2}&-a_{N3}&\cdots&l_{NN}  \end{array} \right] \\
\mathcal{A}_{22}=&\left[\begin{array}{cccc} -a_{21}&-a_{22}&\cdots&-a_{2N} \\ -a_{31}&-a_{32}&\cdots&-a_{3N}\\ \vdots&\vdots& &\vdots  \\ -a_{N1}&-a_{N2}&\cdots&-a_{NN}  \end{array} \right]
\end{align*}

Then $\eqref{19}$ can be rewritten as
\begin{equation}
\label{20}
\dot{\delta}(t)=\Pi \delta(t)+\mathcal{W}e(t)
\end{equation}
where $\Pi=I_{N-1}\otimes A+(\mathcal{L}_{22}+1_{N-1}a_{1}^{*})\otimes BK$, $\mathcal{W}=(\mathcal{A}_{22}+1_{N-1}a_{1}+\mathcal{M})\otimes BK$.

If agent $i$ is triggered, i.e. $f_{i}(t)\geq 0$, then its measurement error $e_{i}(t)$ will be reset to 0. It means that $f_{i}(t)$ will not cross 0 and the measurement $e_{i}(t)$ satisfies $\big\|e_{i}(t)\big\|\leq c_{1}e^{-\alpha t}$ before agent $i$ is triggered. Clearly, $\big\|e(t)\big\|\leq \sqrt{N}c_{1}e^{-\alpha t}$ and $\lim_{t \to \infty}\big\|e(t)\big\|=0$ holds. Therefore, it can be seen that if the matrix $\Pi$ is Hurwitz, then the system $\eqref{20}$ can asymptotically converge to 0 as $t\to \infty$, i.e. the multi-agent system $\eqref{1}$ can achieve consensus under the consensus protocol $\eqref{3}$ and the triggering function $\eqref{5}$.

Following Lemma 3, an invertible matrix can be taken as $S^{-1}=\left[\begin{array}{cc} 1 & 0 \\  -1_{N-1} & I_{N-1} \end{array} \right]$, then it has that $S^{-1}\mathcal{L}S=\left[\begin{array}{cc} 0 & -a_{1}^{*} \\  0 & \mathcal{L}_{22}+1_{N-1}a_{1}^{*} \end{array} \right]$. Therefore, it can be proved as like Lemma 3 that the eigenvalues of the matrix $\Pi$ are the same as the ones of $A+\lambda_{i}(\mathcal{L})BK(i=2,3,\cdots,N)$. As a result, if all the matrices $A+\lambda_{i}(\mathcal{L})BK(i=2,3,\cdots,N)$ are Hurwitz, the matrice $\Pi$ is surely Hurwitz. Then the system $\eqref{20}$ can asymptotically converge to 0 as $t\to \infty$, i.e. the multi-agent system $\eqref{1}$ can achieve consensus under the consensus protocol $\eqref{3}$ and the triggering function $\eqref{5}$.

(Necessity)It is assumed that not all the matrices $A+\lambda_{i}(\mathcal{L})BK(i=2,3,\cdots,N)$ are Hurwitz, so it is clear that the matrice $\Pi$ is not Hurwitz. If the initial value of $\delta(t)$ is not 0, then $\delta(t)$ will go to infinity as $t \to \infty$. So the multi-agent system $\eqref{1}$ can not achieve consensus under the consensus protocol $\eqref{3}$ and the triggering function $\eqref{5}$.

Next, the nonexistence of the Zeno behavior in the control process will be proved. From $\eqref{4}$, it can be derived that
\begin{equation}
\begin{split}
\label{21}
\dot{e}_{i}(t)=&Ae^{A(t-t_{k_{i}}^{i})}x_{i}(t_{k_{i}}^{i})+B\hat{u}_{i}(t) \\
&+\int_{t_{k_{i}}^{i}}^{t}Ae^{A(t-s)}B\hat{u}_{i}(s)ds -\dot{x}_{i}(t)  \\
=&A\big[e^{A(t-t_{k_{i}}^{i})}x_{i}(t_{k_{i}}^{i})+\int_{t_{k_{i}}^{i}}^{t}e^{A(t-s)}B\hat{u}_{i}(s)ds\big]\\
&+B\hat{u}_{i}(t) -Ax_{i}(t)-Bu_{i}(t)  \\
=&Ae_{i}(t)+BK(a_{i}^{*}\otimes I_{n})e^{\Omega_{i}(t-t_{k_{i}}^{i})}\hat{\theta}_{i}(t_{k_{i}}^{i})-Bu_{i}(t)
\end{split}
\end{equation}
And $\eqref{16}$ can be rewritten as
\begin{equation}
\begin{split}
\label{22}
u_{i}(t)=K(l_{i}^{*}\otimes I_{n})\delta(t)+K(l_{i}\otimes I_{n})e(t)
\end{split}
\end{equation}
where $l_{i}^{*}=[l_{i2},l_{i3},\cdots,l_{iN}]$.

Then substituting $\eqref{22}$ into $\eqref{21}$ yields
\begin{equation}
\begin{split}
\label{23}
\dot{e}_{i}(t)=&Ae_{i}(t)+BK(a_{i}^{*}\otimes I_{n})e^{\Omega_{i}(t-t_{k_{i}}^{i})}\hat{\theta}_{i}(t_{k_{i}}^{i}) \\
&-BK(l_{i}^{*}\otimes I_{n})\delta(t)-BK(l_{i}\otimes I_{n})e(t)
\end{split}
\end{equation}
From the triggering function $\eqref{5}$, it can be known that $\big\|e_{i}(t)\big\|\leq c_{1}e^{-\alpha t}$. Then it can be obtained that
\begin{eqnarray}
\label{24}
\lefteqn{\Vert\dot{e}_{i}(t)\Vert } && \nonumber \\
& & \leq \Vert A \Vert \Vert e_{i}(t)\Vert+\Vert BK\Vert \Vert a_{i}^{*}\otimes I_{n}\Vert \Vert e^{\Omega_{i}(t-t_{k_{i}}^{i})}\Vert \Vert \hat{\theta}_{i}(t_{k_{i}}^{i}) \Vert \nonumber  \\
& &+\Vert BK\Vert \Vert l_{i}^{*}\otimes I_{n}\Vert \Vert\delta(t) \Vert+\Vert BK\Vert \Vert  l_{i}\otimes I_{n}\Vert \Vert e(t)\Vert \nonumber \\
& & \leq \Vert A \Vert c_{1}e^{-\alpha t}+\Vert BK \Vert \Vert a_{i}^{*}\otimes I_{n} \Vert  \Vert e^{\Omega_{i}(t-t_{k_{i}}^{i})} \Vert \Vert \hat{\theta}_{i}(t_{k_{i}}^{i}) \Vert \nonumber \\
& &+\Vert BK \Vert \Vert l_{i}^{*}\otimes I_{n} \Vert \Vert \delta(t) \Vert + \Vert BK \Vert \Vert l_{i}\otimes I_{n} \Vert \sqrt{N}c_{1}e^{-\alpha t}\nonumber \\
\end{eqnarray}
From Lemma 3, it can be known that if all the matrices $A+\lambda_{i}(\mathcal{L})BK$ $(i=2,3,\cdots,N)$ are Hurwitz, then all the matrices $\Omega_{i}(i=1,2,\cdots,N)$ and $\Pi$ are also Hurwitz. Then it follows from Lemma 2 that $\Vert e^{\Pi (t-s)} \Vert\leq c_{\Pi}e^{\mu_{\Pi}(t-s)}$ and $\Vert e^{\Omega_{i} (t-t_{k_{i}}^{i})} \Vert\leq c_{\Omega_{i}}e^{\mu_{\Omega_{i}}(t-t_{k_{i}}^{i})}$, where $\mu_{\Omega_{i}}<0, \mu_{\Pi}<0, c_{\Omega_{i}}>0, c_{\Pi}>0$. The solution of $\eqref{20}$ can be obtained.
\begin{equation}
\label{25}
\delta(t)=e^{\Pi t}\delta(0)+\int_{0}^{t}e^{\Pi(t-s)}\mathcal{W}e(s)ds
\end{equation}
According to Lemma 2, it has that
\begin{equation}
\begin{split}
\label{26}
\Vert e^{\Pi(t-s)}\mathcal{W}e(s) \Vert\leq& \beta_{2}e^{\mu_{\Pi}(t-s)}e^{-\alpha s}
\end{split}
\end{equation}
where $\beta_{2}=c_{1}c_{\Pi}\sqrt{N} \Vert \mathcal{W} \Vert$.
So it can be derived that
\begin{equation}
\begin{split}
\label{27}
\Vert \delta(t) \Vert=&\Vert e^{\Pi t}\delta(0)+\int_{0}^{t}e^{\Pi(t-s)}\mathcal{W}e(s)ds \Vert \\
\leq&\beta_{1} e^{\mu_{\Pi}t}+\int_{0}^{t}\beta_{2}e^{\mu_{\Pi}(t-s)}e^{-\alpha s}ds  \\
=&\eta_{1}e^{\mu_{\Pi}t}+\eta_{2}e^{-\alpha t}
\end{split}
\end{equation}
where $\beta_{1}=c_{\Pi}\Vert \delta(0) \Vert$, $\eta_{1}=\beta_{1}+\frac{\beta_{2}}{\vert \mu_{\Pi}+\alpha \vert}$, $\eta_{2}=\frac{\beta_{2}}{\vert \mu_{\Pi}+\alpha \vert}$. Substituting $\eqref{27}$ into $\eqref{24}$ yields
\begin{equation}
\label{28}
\Vert\dot{e}_{i}(t)\Vert\leq \varphi_{i}e^{-\alpha t}+\phi_{i}e^{\mu_{\Pi}t}+\omega_{i}e^{\mu_{\Omega_{i}}(t-t_{k_{i}}^{i})}  \Vert \hat{\theta}_{i}(t_{k_{i}}^{i})\Vert
\end{equation}
where
\begin{eqnarray*}
\varphi_{i}&=&\Vert A \Vert c_{1}+\Vert BK \Vert \Vert l_{i}^{*}\otimes I_{n} \Vert \eta_{2} +\Vert BK \Vert \Vert l_{i}\otimes I_{n} \Vert \sqrt{N}c_{1}  \\
\phi_{i}&=&\Vert BK \Vert \Vert l_{i}^{*}\otimes I_{n} \Vert \eta_{1}  \\
\omega_{i}&=&\Vert BK \Vert \Vert a_{i}^{*}\otimes I_{n} \Vert c_{\Omega_{i}}
\end{eqnarray*}
From the triggering function $\eqref{5}$, it can be seen that when $\Vert \int_{t_{k_{i}}^{i}}^{t}\dot{e}_{i}(s) ds\Vert=c_{1}e^{-\alpha t}$ holds, the events will be triggered.
\\From $\eqref{28}$, it can be known that
\begin{eqnarray}\label{29}
\lefteqn{\Big\Vert \int_{t_{k_{i}}^{i}}^{t}\dot{e}_{i}(s) ds\Big\Vert  \leq  \int_{t_{k_{i}}^{i}}^{t}\Vert\dot{e}_{i}(s)\Vert ds } & &  \hspace{5mm} \nonumber \\
& \leq \int_{t_{k_{i}}^{i}}^{t}(\varphi_{i}e^{-\alpha s}+\phi_{i}e^{\mu_{\Pi}s}+\omega_{i}e^{\mu_{\Omega_{i}}(s-t_{k_{i}}^{i})}\Vert \hat{\theta}_{i}(t_{k_{i}}^{i})\Vert)ds
\end{eqnarray}
where $0<t_{k_{i}}^{i}<t$.

So it can be seen that the event of agent $i$ will not be triggered before $\int_{t_{k_{i}}^{i}}^{t}(\varphi_{i}e^{-\alpha s}+\phi_{i}e^{\mu_{\Pi}s}+\omega_{i}e^{\mu_{\Omega_{i}}(s-t_{k_{i}}^{i})}\Vert \hat{\theta}_{i}(t_{k_{i}}^{i})\Vert)ds=c_{1}e^{-\alpha t}$ holds. Define $t_{1}^{i}$ and $t_{2}^{i}$ are the two neighbouring triggering instants of the agent $i$ satisfying $0<t_{1}^{i}<t_{2}^{i}$. Let $\tau=t_{2}^{i}-t_{1}^{i}$ denote the interval between the two neighbouring triggering instants, and it has been known that $-\alpha, \mu_{\Pi}, \mu_{\Omega_{i}}<0$, so there exists $e^{-\alpha s}$, $e^{\mu_{\Pi} s}$, $e^{\mu_{\Omega_{i}} (s-t_{1}^{i})}$ $\leq 1$. So it has that
\begin{eqnarray}
\label{30}
\lefteqn{\int_{t_{1}^{i}}^{t}(\varphi_{i}e^{-\alpha s}+\phi_{i}e^{\mu_{\Pi}s}+\omega_{i}e^{\mu_{\Omega_{i}}(s-t_{1}^{i})}\Vert \hat{\theta}_{i}(t_{1}^{i})\Vert)ds } & &  \nonumber \\
\leq & \int_{t_{1}^{i}}^{t}(\varphi_{i}+\phi_{i}+\omega_{i}\Vert \hat{\theta}_{i}(t_{1}^{i})\Vert)ds
\end{eqnarray}
And it can be known from the triggering function that $\tau$ is the solution of the function $\Vert \int_{t_{1}^{i}}^{t_{1}^{i}+\tau}\dot{e}_{i}(s) ds\Vert=c_{1}e^{-\alpha (t_{1}^{i}+\tau)}$. Therefore, the value of $\tau$ must be greater than or equal to the solution of the following function, i.e. $\tau \geq \tau^{*}$.
\begin{equation}
\label{31}
 (\varphi_{i}+\phi_{i}+\omega_{i}\Vert \hat{\theta}_{i}(t_{1}^{i})\Vert)\tau^{*}=c_{1}e^{-\alpha (t_{1}^{i}+\tau^{*})}
\end{equation}
Thus there must be a positive lower bound on interval between any two neighbouring event-triggered instants. The Zeno behavior is proved nonexistent. The proof is completed. 
\end{Proof}

For the linear continuous-time multi-agent system $\eqref{1}$ with the event-triggered consensus protocol $\eqref{3}$ and the triggering function $\eqref{5}$, an appropriate $K$ can be chosen to ensure that all the matrices $A+\lambda_{i}(\mathcal{L})BK$ $(i=2,3,\cdots,N)$ are Hurwitz by the following steps.
\begin{Step}
It has been assumed that $(A,B)$ in $\eqref{1}$ is stabilizable in Assumption 1,thus the Riccati equation $A^{T}P+PA-PBB^{T}P+I_{n}=0$ has a unique nonnegative definite solution $P$, and all the eigenvalues of $A-BB^{T}P$ are in the open left half plane\cite{ref25}.
\end{Step}
\begin{Step}
For any $\sigma \geq 1$ and $\omega\in R$, all the eigenvalues of $A-(\sigma+j\omega)BB^{T}P$ $(j^{2}=-1)$ are in the open left half plane\cite{ref26}.
\end{Step}
\begin{Step}
Select $K=-cB^{T}P$, where $c>\frac{1}{min(Re(\lambda_{i}(\mathcal{L})))}$, where $\lambda_{i}(\mathcal{L})\neq 0 \quad(i=2,3,\cdots,N)$.
\end{Step}

\section{Simulation}
In this section, a numerical examples is given to illustrate the effectiveness and the advantage of the method proposed in this paper.
\\Consider a linear continuous-time multi-agent system consists of six agents. The dynamics model of agent $i$ is described by the system $\eqref{1}$ with
\[
A=\left[ {\begin{array}{*{20}c}
   { 0} & ~1  \\
   -1 & { ~0}  \\
\end{array}} \right],~~ B  = \left[ {\begin{array}{*{20}c}
   1 \\
  1  \\
\end{array}} \right]
\]
The communication topology among the six agents is described by a weighted graph as shown in Figure 1. Let the initial state of the system be $x_{1}(0)=[0.4~~0.3]^T$, $x_{2}(0)=[0.5~~0.2]^T$, $x_{3}(0)=[0.6~~0.1]^T$, $x_{4}(0)=[0.7~~0]^T$, $x_{5}(0)=[0.8~~-0.1]^T$, $x_{6}(0)=[0.4~~-0.2]^T$. The feedback gain matrix is designed as $K=[-2.2~~-1.1]$. And the other parameters are $c_{1}=0.6$ and $\alpha=0.4$. The Laplacian matrix of the weighted graph is
\[
L=\left[ {\begin{array}{*{20}c}
   3 & ~0 & ~0 & ~-1 & ~-1& ~-1  \\
  -1& ~1& ~0& ~0& ~0& ~0\\
   -1& ~-1& ~2& ~0& ~0& ~0\\
   -1& ~0& ~0& ~1& ~0& ~0\\
   0& ~0& ~0& ~-1& ~1& ~0\\
   0& ~0& ~0& ~0& ~-1& ~1
\end{array}} \right]
\]

\begin{figure}[H]
\centering\includegraphics[scale=0.4]{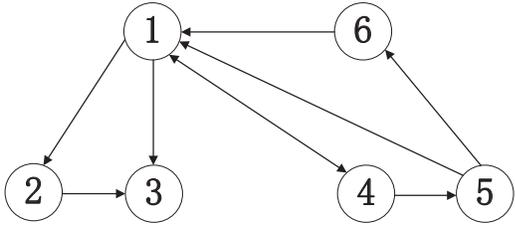}
\caption{Communication topology among the six agents}
\label{fig1}
\end{figure}

Fig 2 shows the state trajectories of all the six agents. It can be seen that the linear continuous-time multi-agent system can achieve consensus, which means the event-triggered consensus protocol proposed in this paper can solve the consensus problem of multi-agent systems effectively. In Fig 3, the measurement error of each agent and the threshold of errors are presented. It can be seen that when the measurement error reaches the threshold, the event is triggered, then the measurement error is reset to zero.

\begin{figure}[htp]
\centering\includegraphics[width=\hsize]{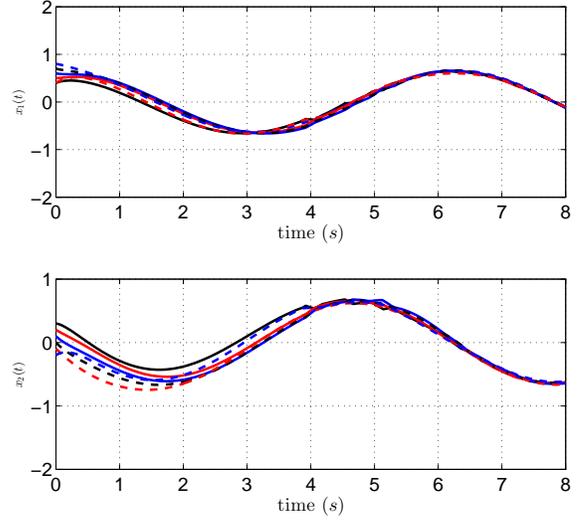}
\caption{State trajectories of all the agents}
\label{fig2}
\end{figure}

\begin{figure}[htp]
\centering\includegraphics[width=\hsize]{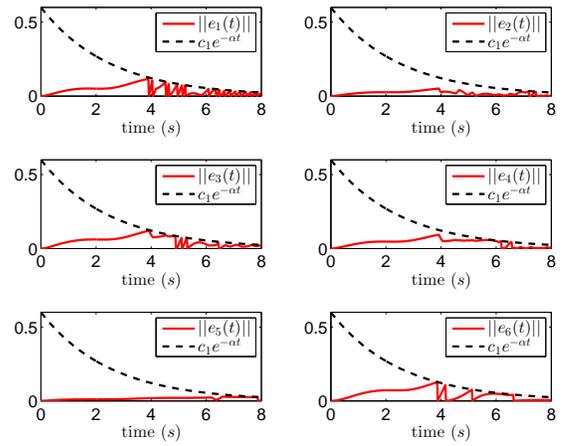}
\caption{Measurement errors and the threshold of errors by using the method in this paper}
\label{fig3}
\end{figure}


Table 1 lists comparisons between the methods in \cite{ref20} \cite{ref21} and the method this paper in terms of event-triggered times of each agent. It can be seen that the event-triggered times using the method in this paper is much fewer than using the methods in \cite{ref20} and \cite{ref21}.

\begin{table}[htp]
\caption{The event-triggered times using different methods}
\centering
\begin{tabular}{cc c c c }
\hline
\centering agent  & \cite{ref20}  &\cite{ref21}& our method \\
\hline
\centering 1 &25& 38   & 18 \\
\hline
\centering 2  &12& 16  & 3 \\
\hline
\centering 3  &19& 24  & 5 \\
\hline
\centering 4  &12& 11  & 3 \\
\hline
\centering 5  &12& 12  & 2 \\
\hline
\centering 6  &11& 16  & 5 \\
\hline
\end{tabular}
\end{table}

\section{Conclusion}

This paper has investigated the event-triggered consensus for linear continuous-time multi-agent systems under the directed communication topology based on a predictor. A new event-triggered protocol has been designed based on a state predictor for the linear continuous-time multi-agent systems to achieve consensus without continuous communication. The consensus protocol provided in this paper only requires each agent to monitor its state to determine the event-triggered instants. And the Zeno behavior has been proved to be nonexistent. On the other hand, an advantage of the method proposed in this paper is that it can make the multi-agent systems achieve consensus with much fewer event-triggered times. So the method in this paper can reduce the unnecessary communication among agents more effectively and save more communication costs.

\flushend

\begin{thebibliography}{0}
\bibitem{ref1}
Maes, P. (1990). Desiging autonomous agents: Theory and practice from biology to engineering and back. Amsterdam, Netherlands: Elsevier Science.

\bibitem{ref2}
Olfati-saber, R., \& Murray, R. M. (2004). Consensus problems in networks of agents with switching topology and time-delays. {\it IEEE Transactions on Automatic Control}, 49(9), 1520-1533.

\bibitem{ref3}
Hong, Y., Hu, J., \& Gao, L. (2006). Tracking control for multi-agent consensus with an active leader and variable topology. {\it Automatica}, 42(7), 1177-1182.

\bibitem{ref4}
Yu, W., Chen, G., \& Cao, M. (2010). Some necessary and sufficient conditions for second-order consensus in multi-agent dynamical systems. {\it Automatica}, 46(6), 1089-1095.

\bibitem{ref5}
Song, Q., Cao, J., \& Yu, W. (2010). Second-order leader-following consensus of nonlinear multi-agent systems via pinning control. {\it Systems \& Control Letters}, 59(9), 553-562.

\bibitem{ref6}
He, W., \& Cao, J. (2011). Consensus control for high-order multi-agent systems. {\it IET Control Theory \& Applications}, 5(1),231-238.

\bibitem{ref7}
Zhou, B., \& Lin, Z. (2014). Consensus of high-order multi-agent systems with large input and communication delays. {\it Automatica}, 50(2), 452-464.


\bibitem{ref8}
Jiang, F., Xie, G., Wang, L., \& Chen, X. (2008). The $\chi$-consensus problem of high-order multi-agent systems with fixed and switching topologies. {\it Asian Journal of Control}, 10(2), 246-253.

\bibitem{ref9}
Ma, C. Q., \& Zhang, J. F. (2010). Necessary and sufficient conditions for consensusability of linear multi-agent systems. {\it IEEE Transactions on Automatic Control}, 55(5), 1263-1268.

\bibitem{ref10}
Gu, G., Marinovici, L., \& Lewis, F. L. (2012). Consensusability of discrete-time dynamic multiagent systems. {\it IEEE Transactions on Automatic Control}, 57(8), 2085-2089.

\bibitem{ref11}
Cao, Y. C., Yu, W. W., Ren, W., \& Chen, G. R. (2013). An Overview of Recent Progress in the Study of Distributed Multi-Agent Coordination. {\it IEEE Transactions on Industrial Informatics}, 9(1), 427-438.

\bibitem{ref12}
Kwang-Kyo, O., Myoung-Chul, P., \& Hyo-Sung, A. (2015). A survey of multi-agent formation control. {\it Automatica}, 53, 424-440.

\bibitem{ref13}
Huang, N., Duan, Z., \& Zhao, Y. (2014). Leader-following consensus of second-order nonlinear multi-agent systems with directed intermittent communication. {\it IET Control Theory \& Applications}, 8(10), 782-795.

\bibitem{ref14}
Fan, Y., Feng, G., Wang, Y., \& Song, C. (2013). Distributed event-triggered control of multi-agent systems with combinational measurements. {\it Automatica}, 49(2), 671-675.

\bibitem{ref15}
Cheng, L., Wang, Y., Hou, Z. G., Tan, M., \& Cao, Z. (2013). Sampled-data based average consensus of second-order integral multi-agent systems: switching topologies and communication noises. {\it Automatica}, 49(5), 1458-1464.

\bibitem{ref16}
Zhang, H., Feng, G., Yan, H. C.,\& Chen, Q. J. (2014). Consensus of multi-agent systems with linear dynamics using event-triggered control. {\it IET Control Theory and Applications}, 8(18), 2275-2281.

\bibitem{ref17}
Dimarogonas, D. V., Frazzoli, E., \& Johansson, K. H. (2012). Distributed eventtriggered control for multi-agent systems. {\it IEEE Transactions on Automatic Control}, 57(5), 1291-1297.

\bibitem{ref18}
Seyboth, G. S., Dimarogonas, D. V., \& Johansson, K. H. (2013). Event-based broadcasting for multi-agent average consensus. {\it Automatica}, 49(1), 245-252.

\bibitem{ref19}
Zhang, H., Feng, G., Yan, H. C., \& Chen, Q. J. (2014). Observer-Based Output Feedback Event-Triggered Control for Consensus of Multi-Agent Systems. {\it IEEE Transactions on Industrial Electronics}, 61(9), 4885-4894.

\bibitem{ref20}
Zhu, W., Jiang, Z. P., \& Feng, G. (2014). Event-based consensus of multi-agent systems with general linear models. {\it Automatica}, 50(2), 552-558.

\bibitem{ref21}
Yang, D. P., Ren, W., Liu, X. D., \& Chen, W. S. (2016). Decentralized event-triggered consensus for linear multi-agent systems under general directed graphs. {\it Automatica}, 69, 242-249.

\bibitem{ref22a}
Sun, J., Chen, J., \& Liu, G.P. (2011). State feedback controller design and stability analysis of networked predictive control systems. In {\it the 2nd International Conference on Intelligent Control and Information Processing} (785-790).

\bibitem{ref22b}
Zhou, B. (2014). Pseudo-predictor feedback stabilization of linear systems with time-varying input delays. {\it Automatica}, 50, 2861-2871.

\bibitem{ref22c}
Cacace, F., Germania, A. \& Manesb, C. (2014) Exponential stabilization of linear systems with time-varying delayed state feedback via partial spectrum assignment.
{\it Systems \& Control Letters}, 69, 47-52.


\bibitem{ref22}
Ren, W., \& Cao, Y. (2011). Distributed Coordination of Multi-agent Networks: Emergent Problems, Models, and Issues. London:Springer-Verlag.

\bibitem{ref23}
Loan, C. V. (1977). The sensitivity of the matrix exponential. {\it Siam Journal on Numerical Analysis}, 14(6), 971-981.

\bibitem{ref24}
Brewer, J. (1978). Kronecker products and matrix calculus in system theory. {\it IEEE Transactions on Circuits \& Systems}, 25(9), 772-781.

\bibitem{ref25}
Cheng, Z., \& Ma, S. (2006). Linear System Theory. Beijing, China: Science Press.

\bibitem{ref26}
Ma, C. Q. (2009). System Analysis and Control Synthesis of Linear MultiAgent Systems. Ph.D. dissertation, Academy of Mathematics and Systems Science, Chinese Academy of Sciences, Beijing, China.
\end{thebibliography}
\end{document}